\documentclass[preprint, superscriptaddress]{revtex4-2}
\usepackage{graphicx}
\usepackage{longtable}
\usepackage{array, makecell}
\usepackage{tabularx}

\usepackage{mathtools}
\usepackage[colorlinks = true,linkcolor = blue,urlcolor  = blue,citecolor = blue,anchorcolor = blue]{hyperref}
\usepackage{xcolor, soul}
\usepackage{color}

\usepackage{makecell}

\DeclarePairedDelimiter\ket{\lvert}{\rangle}

\begin{document}


\title{Coherent electric field manipulation of nuclear spin qudit}


\author{Sumin Lim}
\thanks{These three authors contributed equally}
\affiliation{CAESR, Department of Physics, University of Oxford, The Clarendon Laboratory, Parks Road, Oxford OX1 3PU, UK}
\affiliation{Graduate School of Quantum Science and Technology, KAIST, Daejeon 34141, Republic of Korea}

\author{Mikhail V. Vaganov}
\thanks{These three authors contributed equally}
\affiliation{CAESR, Department of Physics, University of Oxford, The Clarendon Laboratory, Parks Road, Oxford OX1 3PU, UK}

\author{Niccol\'{o} Fontana}
\thanks{These three authors contributed equally}
\affiliation{CAESR, Department of Physics, University of Oxford, The Clarendon Laboratory, Parks Road, Oxford OX1 3PU, UK}

\author{Junjie Liu}
\email{junjie.liu@qmul.ac.uk}
\affiliation{School of Physical and Chemical Sciences, Queen Mary University of London, London E1 4NS, UK}

\author{Arzhang Ardavan}
\email{arzhang.ardavan@physics.ox.ac.uk}
\affiliation{CAESR, Department of Physics, University of Oxford, The Clarendon Laboratory, Parks Road, Oxford OX1 3PU, UK}

\begin{abstract}
Spins in condensed matter, especially well-isolated nuclear spins, offer attractive quantum degrees of freedom for computing, sensing, and networking because of their long coherence times. The possibility of electric-field control is an important feature for practical scalable quantum technologies, but, typically, nuclear spins couple only weakly to electric fields in conventional semiconductor hosts, limiting operation efficiency. Here we show that a choice of a highly polarizable oxide host can overcome this bottleneck. In Mn$^{2+}$ doped ZnO, electric-field modulation of the spin Hamiltonian is amplified by hyperfine-coupled electron spins, and offers efficient electric-field manipulation of an $I=5/2$ nuclear spin qudit, in a manner analogous to the hyperfine enhancement of conventional nuclear magnetic resonance. We demonstrate both resonant and non-resonant coherent manipulation using a single uniaxial electric field applied along the crystalline c-axis, the polarization axis of ZnO. This approach allows universal single-qudit gate operations with efficiencies comparable to or exceeding those of conventional magnetic-field driving. These results support the deployment of doped oxides as active host materials for electrically controllable spin qubits, highlighting the importance of materials design in developing scalable quantum technologies.
\end{abstract}


\maketitle

Among various physical modalities for qubits, spins in condensed matter materials~\cite{Burkard2023}, such as electron and nuclear spins of donors in silicon~\cite{Kane1998,Morton2008,Pla2012a,Asaad2019,Mdzik2022,Gilbert2023,Reiner2024,Stemp2025,Yu2025}, have emerged as promising candidates for solid-state quantum computing due to their long spin coherence time and the possibility of scaling up from a single qubit to larger scale quantum processors. Furthermore, high-spin nuclei have been identified as spin-based qudits that offer a hardware-efficient approach for the implementation of various quantum algorithms and quantum error correction~\cite{Chiesa2020,Gross2021,Lim2023,Lim2025}. 

Conventional nuclear magnetic resonance (NMR) relies on a static magnetic field ($B_0$) to tune the energy levels of nuclear spins and a radio-frequency (RF) magnetic control field ($B_1$) to induce a transition between discrete Zeeman states. However, it is challenging to confine magnetic fields at the nanometre scale, preventing selective control of individual spins on an architecturally relevant scale. Electric fields, by contrast, can be efficiently routed and confined within complex nanoscale devices, making electric-field control the preferred approach at nanometre scales\cite{Laucht2015,Asaad2019,Gilbert2023}. However, spin control by this method relies on nuclear spin-electric couplings (nSECs), which, in silicon, rely heavily on techniques such as strain engineering~\cite{Bradbury2007,Lo2014,Pica2014}, and are typically very weak. As a result, very strong electric fields, $\sim 10^6$ V/m, are needed to achieve single-spin operation times in the millisecond range. While such long operations can be tolerated in certain circumstances, e.g., spins doped in isotopically purified $^{28}$Si, thanks to their impressive coherence times~\cite{Muhonen2014}, it puts stringent requirements on the operation conditions (such as temperature, environmental nuclear bath), under which electric nuclear spin control would be beneficial. Furthermore, such slow spin control leads to limitations for applications such as quantum sensing~\cite{Degen2017}, where long control pulse durations impact directly on the sensor performance. Therefore, we are motivated to explore alternative materials exhibiting enhanced nSECs that might allow fast spin manipulation. 

Oxides offer interesting advantages as material platforms for hosting qubits while also leveraging the mature infrastructure developed for microelectronics. In particular, complex oxides that include transition metal ions can exhibit interactions between charge, spin, orbital and lattice degrees of freedom, enabling emergent functionalities relevant to next-generation logic and memory devices~\cite{Manipatruni2019,Spaldin2019}. For spin-based qubits, the large dielectric constant and the lack of inversion symmetry in piezo-/ferro-electrics oxides suggest that crystal field terms in the spin Hamiltonian can be sensitive to external electric fields, supporting strong electron spin-electric couplings (eSECs) for rapid spin manipulations~\cite{George2013,Liu2021,Fang2022,Liu2024}. However, analogous interactions involving the nuclear spins of magnetic defects in oxides remain largely unexplored.

In this work, we investigate coherent nSECs of Mn$^{2+}$ defects in a piezoelectric single crystal zinc oxide (ZnO). We find that the nSECs associated with the Mn$^{2+}$ nuclear spin are substantially stronger than those measured in donor spins in silicon~\cite{Asaad2019}, enabling, in principle, faster electrical control of both the nuclear spin energy and quantum coherence. Our analysis shows that the observed nSECs are enhanced by the hyperfine-coupled electron spin, with the eSECs providing substantial, and in some cases dominant, contributions to the nSECs.

Mn$^{2+}$ (atomic configuration [Ar]3d$^5$4s$^0$) substituting Zn in single-crystal ZnO offers an electron spin $S = 5/2$  and nuclear spin $I = 5/2$ hyperfine-coupled system with a well-defined 36-dimensional Hilbert space. 
The low temperature spin spectrum of Mn$^{2+}$ in ZnO can be described by the  Hamiltonian~\cite{Hausmann1968,Bttcher2014}
\begin{equation}
H = \mu_B g_e \hat{\mathbf{S}}\cdot\mathbf{B}_0 + \mu_N g_n \hat{\mathbf{I}}\cdot\mathbf{B}_0 + A_\parallel\hat{S}_z\hat{I}_z + A_\perp(\hat{S}_x\hat{I}_x + \hat{S}_y\hat{I}_y) + D \hat{S_z}^2 + F\hat{S_z}^4 + Q_{zz}\hat{I}^2_z
\label{H2}
\end{equation}
where $D$ and $F$ describe the magnetic anisotropy of the electron spin, $A_\parallel$ and $A_\perp$ correspond to the axial and transverse components of the hyperfine interaction, respectively, and $Q_{zz}$ represents the axial nuclear quadrupole interaction.

The nuclear spin coherence time ($T_{2n}$), limited by the electron spin-lattice relaxation time ($T_{1e}$), exceeds 100~ms below 4~K. Measurement of the coherent eSEC reveal that the electron spin zero-field splitting (ZFS) parameter $D$ is modulated by electric fields (E-fields) applied along the crystallographic $c$-axis of the ZnO host~\cite{George2013}. 

\begin{figure}
\includegraphics[width=16cm]{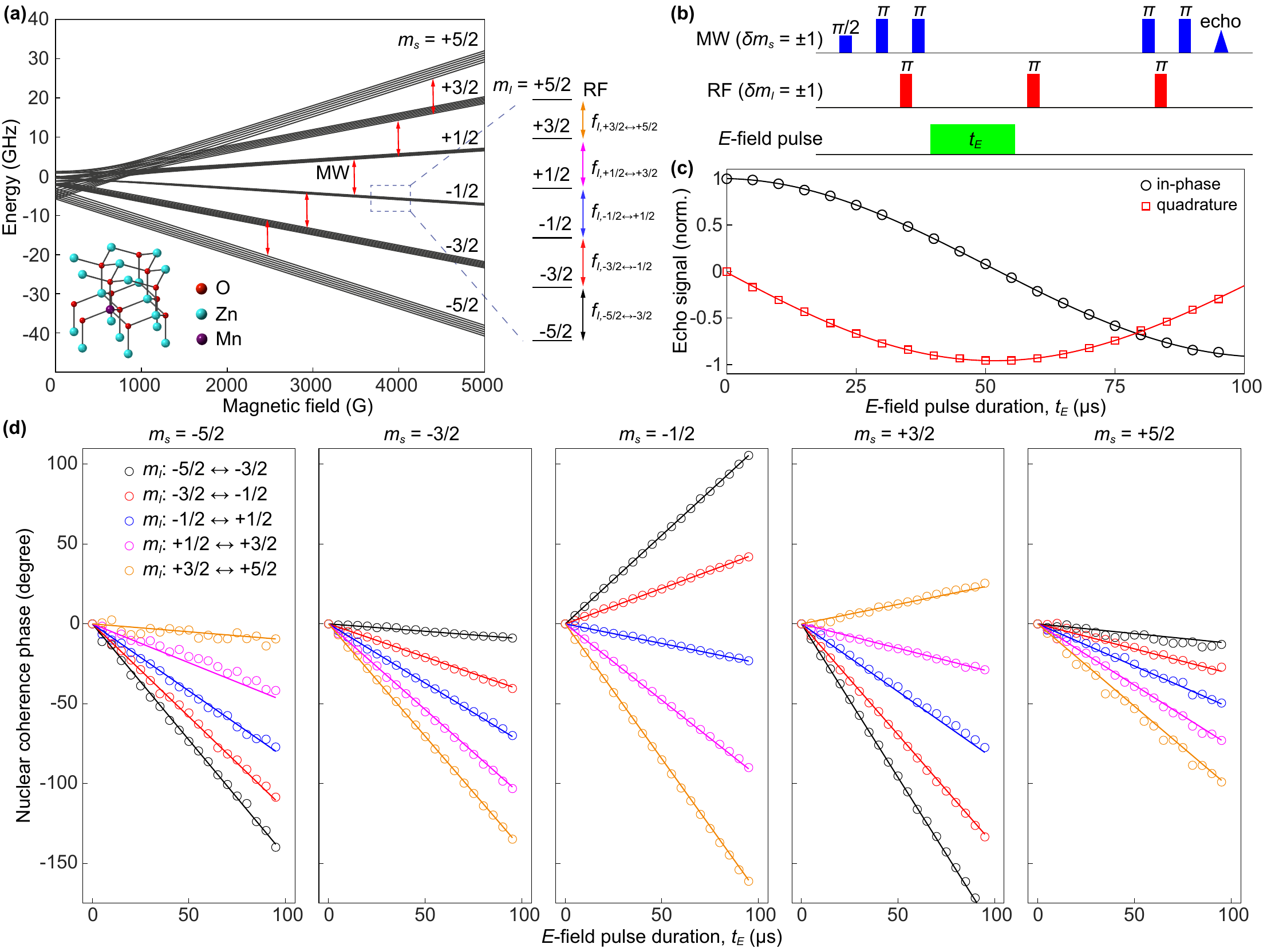}
\caption{\textbf{nSECs for Mn$^{2+}$ defects in ZnO.} \textbf{(a)} Left: the spin energy levels of Mn$^{2+}$ defects in ZnO showing the splitting between different electron spin multiplets labeled by $m_s$. Electron spin coherence between any adjacent electron spin states can be generated using X-band microwave (MW) pulses in the experimental magnetic field range, as indicated by the red arrows. Right: close-up of the $m_s = -1/2$ multiplet showing the nuclear spin energy levels and RF transitions between them. \textbf{(b)} The modified ENCT sequence used to measure nSECs. The phase of the final echo is recorded as a function of the duration or the amplitude of the E-field pulse. \textbf{(c)} Representative SEC data recorded for the $\ket{m_s = -1/2, m_I = +3/2} \leftrightarrow \ket{m_s = -1/2, m_I = +5/2}$ transition. The in-phase and quadrature components of the echo vary as cosine and sine functions (solid lines) as a function of $t_E$ respectively, revealing the E-field-induced modulation of the observed echo phase, $\delta \varphi_E$. \textbf{(d)} The E-field-induced nuclear coherence phase shift recorded for different transitions. The solid lines represent the best fit to the data with the SEC parameters in the main text. 
}
\label{Schematic}
\end{figure}

When a sufficiently large static magnetic field $\mathbf{B}_0$ is applied parallel to the crystallographic $c$-axis (i.e. $\mathbf{B}_0 \parallel z$) as shown in Fig.~\ref{Schematic}\textbf{(a)}, the eigenstates of Mn$^{2+}$ can be (approximately) presented as $\ket{m_s, m_I}$, where $m_s$ and $m_I$ are the electron and nuclear spin projections on the $\mathbf{B}_0$ direction, respectively. However, in the magnetic field range studied here, the transverse component of the hyperfine interaction, $A_\perp(\hat{S}_x\hat{I}_x + \hat{S}_y\hat{I}_y)$, mixes the electron and nuclear spin states, leading to a significant enhancement of the interaction between the nuclear spin and applied magnetic fields. Importantly, the combination of the nuclear quadrupole interaction, and second order shifts from the hyperfine interaction lead to spectrally distinguishable nuclear spin transitions. Thus, for each $m_s$ projection, five non-degenerate nuclear spin $\ket{m_I} \leftrightarrow \ket{m_I \pm 1}$ transitions can be observed via electron-nuclear double resonance (ENDOR) experiments.

The nSEC for the Mn$^{2+}$ defects is measured employing a modified electron-nuclear coherence transfer (ENCT) sequence as shown in Fig.~\ref{Schematic}\textbf{(b)}~\cite{Morton2008}. First, a spin coherence is generated within the electron spin manifold and subsequently transferred to the nuclear spin manifold. An E-field pulse is then applied while the quantum coherence is stored in the nuclear spin subspace. This E-field pulse modulates the nuclear spin transition frequency, thereby inducing a phase shift in the nuclear spin coherence, which is subsequently transferred back to the electron spin manifold for detection. 

Fig.~\ref{Schematic}\textbf{(c)} shows representative data depicting the phase shift of the measured echo as a function of the E-field pulse duration. The phase of the final echo signal evolves as a function of the duration of the applied E-field, demonstrating modulation of the nuclear quantum phase by the applied E-field. The magnitude and polarity of the nSEC can be extracted from the relation $\delta\varphi_E = \delta f_E t_E$, where $\delta\varphi_E$ is the E-field-induced phase shift, $\delta f_E$ is the E-field-induced shift of the nuclear spin transition frequency, and $t_E$ is the duration of the applied E-field pulse.

We performed nSEC experiments with all $m_s$ manifolds except $m_s = +1/2$, where the nuclear spin transition rates are small due to an accidental suppression of second order hyperfine effects in this magnetic field range~\cite{Bttcher2014}. The results are summarised in Fig.~\ref{Schematic}\textbf{(d)}. The observed nSECs depend on both $m_s$ and $\Delta m_I = \pm1$ transition, suggesting that both the hyperfine and the nuclear quadrupole interactions are modulated by the applied E-field.

 By fitting all the SEC data simultaneously, we found it is necessary to include E-field modulations of $A_\parallel$ and $Q_{zz}$ parameters, together with the previously reported E-field modulation of the electron spin ZFS, to obtain good fits to the experimental data. The ENDOR frequency is less sensitive to modulations to the transverse components $A_\perp$ in Eqn.~\ref{H2}, though the inclusion of a SEC for $A_\perp$ lead to a significant improvement to the fit. This is consistent with the axial piezoelectric nature of the host ZnO matrix, where the electric dipole of the lattice is aligned with the crystallographic $c$-axis: an E-field applied in this direction couples strongly to the electric dipole and modulates the crystal structure, leading to the E-field sensitivity for the axial parameters in Eqn.~\ref{H2}. 

\begin{table}[t]
\centering
\caption{\textbf{The E-field-sensitive Hamiltonian parameters and the corresponding E-field sensitivity.} The E-field sensitivity of each parameter, defined as the E-field-induced modulation of the parameter per V/m, is obtained by fitting the data shown in Fig.~\ref{Schematic}\textbf{(d)}. The individual contribution to the nSECs for representative nuclear spin transitions (at  $E = 4\times10^5$ V/m) are also included. Note the effect of varying $D$ and $F$ are indistinguishable. Therefore, we only included the E-field-induced modulation of $D$ in fitting the data.}
\label{Zgateparameter}
\begin{tabular}{|c|c|c|c|c|}
\hline
Hamiltonian term & D & $A_\parallel$ & $A_\perp$ & $Q_{zz}$ \\
\hline
$E = 0$ value (MHz) & -707  & -225 & -225 & 0.32 \\
\hline
E-field sensitivity [kHz/(kV/cm)] & \quad 77 $\pm 2 \quad $ & $ 0.103 \pm 0.003$ & $-0.10 \pm 0.02$ & $-0.104 \pm 0.003$ \\
\hline
\hline
\multicolumn{5} {|l|}{\hspace{0.5cm}Contributions to nSECs for selected transitions (kHz)}\\
\hline
$m_s=-1/2, m_I: -5/2 \leftrightarrow -3/2$ & 1.63 & -0.21  & -0.27 & 1.75  \\
\hline
$m_s=-1/2, m_I: -1/2 \leftrightarrow +1/2$ & -0.334 & -0.203  & -0.208  & 0 \\
\hline
\end{tabular}
\end{table}

The observed nSECs are strong compared with previously reported values, particularly for dopants in silicon ~\cite{Bradbury2007,Lo2014,Pica2014,Asaad2019}, perhaps owing to the piezoelectric host matrix. 
Furthermore, our analysis shows that the mixing between nuclear and electron spin states allows the relatively large eSECs to make a substantial, and sometimes dominant, contribution to the nSECs. 
This mechanism is analogous to the hyperfine enhancement effect~\cite{Sangtawesin2014,Sangtawesin2016} in conventional NMR experiments, where nuclear spin transition rates can be enhanced through mixing with electron spin states associated with large gyromagnetic ratios. In particular, for the nuclear spin transition between the $m_I = \pm 1/2$ states, which is insensitive to $Q_{zz}$, the coupling to the electron spin offers the only mechanism for nSEC. The fact that multiple terms in the spin Hamiltonian are modulated by the E-field provides a range of pathways for electrically manipulating nuclear spin transitions that might be insensitive to particular terms.

The nSECs for axial parameters allow efficient implementation of phase gates ($Z$-gates) on nuclear spin coherences, whether the coherence involves just two states (i.e.\ a qubit) or multiple states (qu\textit{d}it). Using the nSECs described above, it is possible to perform selective $Z$-gates on arbitrary superpositions of qu\textit{d}it levels (see the Supplementary Information for experimental confirmation of our conditional control on target qu\textit{d}it subspaces). This operation constitutes an essential component of a universal gate set for qu\textit{d}its~\cite{Brennen2005,Wang2020}, and is a key element of the qudit version of Quantum Fourier Transform~\cite{Chi2022,Rubn-Osanz2026}. These results demonstrate that E-field manipulation on nuclear spins can be extended to general qu\textit{d}it-based quantum computing.

The nSECs that we characterise above can also be exploited to drive nuclear electric-dipole spin resonances (NERs) using high-frequency (AC) resonant E-fields applied along the same crystallographic $c$-axis. 
The AC electric field was applied to the $c$-axis of the ZnO crystal using the same electrodes employed for the DC measurements. We chose to use this parallel plate geometry, rather than using an impedance-matched RF antenna, to guarantee a good separation between the electric and magnetic components of the AC field.
This configuration suppresses parasitic AC magnetic fields at the sample thereby distinguishing reliably between electric-dipole and magnetic-diople transitions, albeit at the cost of reduced RF power, and therefore E-field amplitude, owing to the impedance mismatch at the experimental RF frequencies we use. 

By applying the AC RF E-field through the same electrodes, we demonstrate that both resonant and non-resonant nuclear spin phase control, i.e. $X$ ($Y$) and $Z$ gates, respectively, can be implemented with electric fields along the crystal $c$-axis. This approach enables a single physical control gate driving all spin operations, offering significant advantages for future device architectures.

\begin{figure}
\includegraphics[width=14cm]{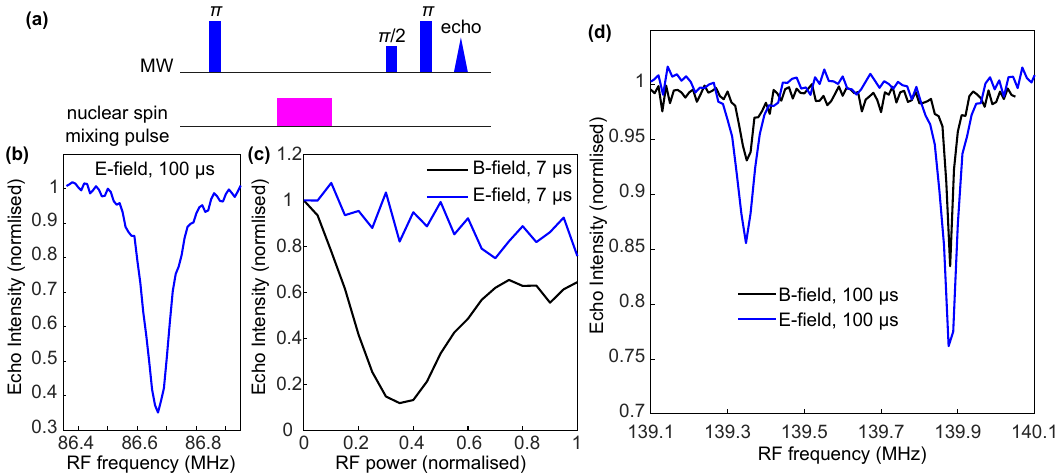}
\caption{\textbf{NERs with $\mathbf{B_0} \parallel$ \textit{c}-axis.} \textbf{(a)} The modified Davies ENDOR pulse sequence for investigating NERs. The RF nuclear spin transition pulse (indicated by the purple block) can be either an AC B-field or an AC E-field pulse. \textbf{(b)} The NER ENDOR spectrum recorded with an 100~$\mu$s E-field RF pulse. The resonance corresponds to the $\ket{m_s = -1/2, m_I = -1/2} \leftrightarrow \ket{m_s = -1/2, m_I = +1/2}$ transition. \textbf{(c)} Nutation experiments performed with a B-field (black) or an E-field (blue) RF pulse. The pulse width is 7~$\mu$s in both cases. \textbf{(d)} The B-field and E-field ENDOR spectra recorded within the $m_s = +1/2$ electron spin multiplet. The two resonances correspond to the $\ket{m_s = +1/2, m_I = -3/2} \leftrightarrow \ket{m_s = +1/2, m_I = -1/2}$ (left) and $\ket{m_s = +1/2, m_I = -1/2} \leftrightarrow \ket{m_s = +1/2, m_I = +1/2}$ (right) transitions.}
\label{AC_E_Davies}
\end{figure}

We demonstrate the excitation of NERs using the modified ENDOR sequence shown in Fig.~\ref{AC_E_Davies}(a). A selective nuclear spin mixing pulse, which can be either a magnetic field (B-field) or an E-field RF pulse, modifies the populations of the spin states and consequently alters the final echo signal. We perform experiments with both B-field and E-field pulses so that we may directly compare standard NMRs with NERs. Figure~\ref{AC_E_Davies}(b) shows a representative NER spectrum recorded within the $m_s = -1/2$ electron spin manifold. When the frequency of the applied E-field pulse matches the transition frequency between the $\ket{m_s = -1/2, m_I = -1/2}$ and $\ket{m_s = -1/2, m_I = +1/2}$ nuclear spin states, the E-field pulse drives population transfer between the two states, resulting in a change in the final electron spin echo intensity. The observed resonance frequency agrees with that obtained from conventional (B-field) ENDOR measurements, confirming the assignment of the transition.

This result confirms that nuclear electric resonance (NER) can be driven using an RF E-field, although the resulting nutation rate is lower than that achieved with RF B-field pulses, as shown in Fig.~\ref{AC_E_Davies}(c). In our experiments, the RF B-field generated by the ENDOR coil is approximately 1~$\mathrm{G}/\sqrt{\mathrm{W}}$, where W is the RF pulse power. At the maximum applied power [corresponding to the normalised RF power of 1 in Fig.~\ref{AC_E_Davies}(c)], we estimate RF field amplitudes of approximately 10 G for the B-field and $7\times10^4~\mathrm{V/m}$ for the E-field.

For nuclear spin transitions within the $m_s=-1/2$ electron spin manifold, the nutation rates of conventional NMR are substantially higher than those of NER driven at the same RF power, owing to the hyperfine enhancement effect that amplifies the magnetic-dipole transition rates. In contrast, for transitions within the $m_s=+1/2$ electron spin manifold, the hyperfine enhancement is accidentally suppressed, making the RF E-field comparatively more effective at driving the transitions. Consequently, the E-field-driven ENDOR intensities are larger in this manifold, as shown in Fig.~\ref{AC_E_Davies}(d).

\begin{figure}
\includegraphics[width=16cm]{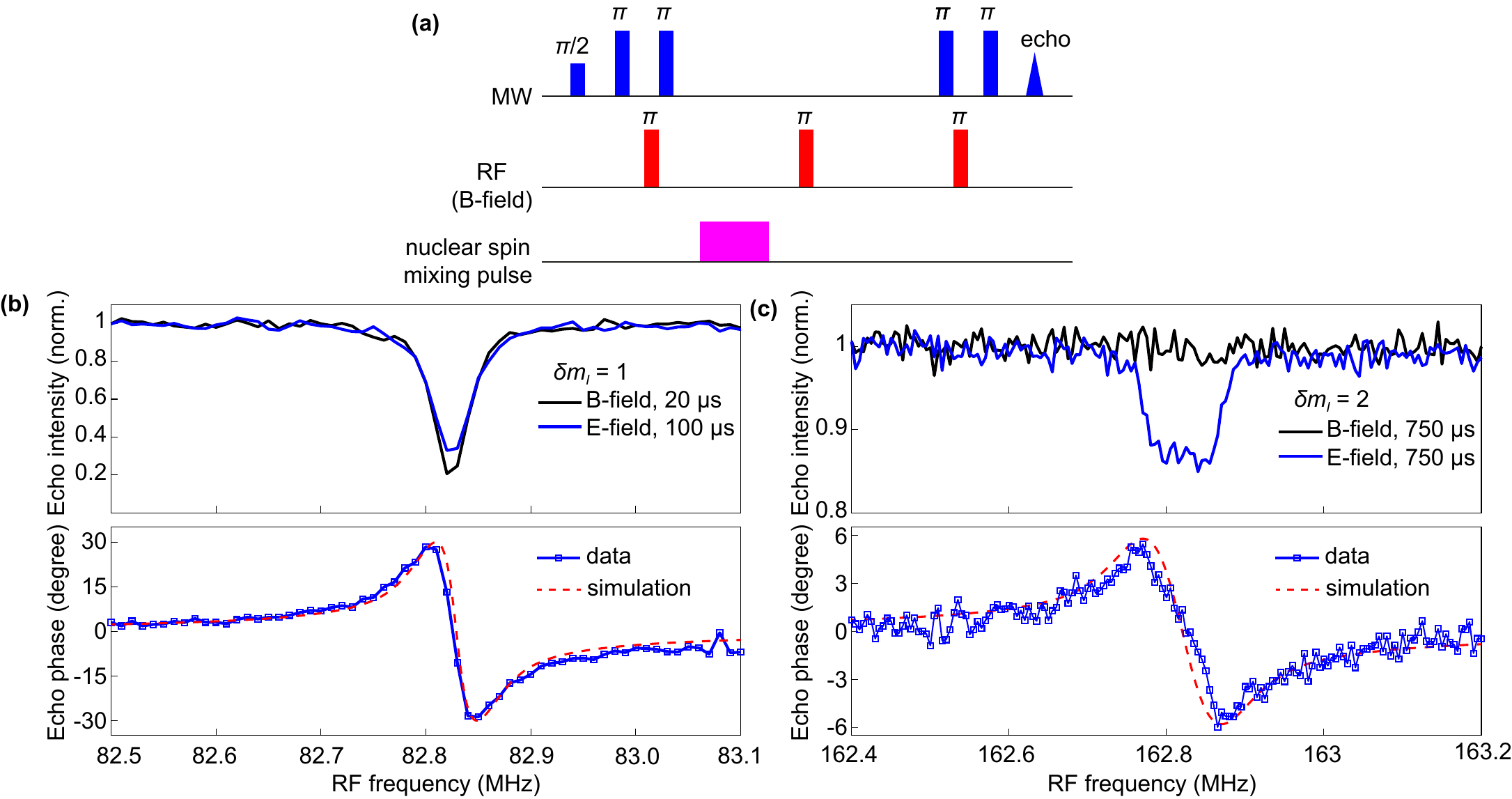}
\caption{\textbf{NERs with $\mathbf{B_0} \perp$ \textit{c}-axis.} \textbf{(a)} The electron-nuclear coherence transfer sequence employed for the investigation of NER. The echo intensity recorded with conventional magnetic (black) and electric (blue) RF pulses are shown in the top panels of \textbf{(b)} and \textbf{(c)} for $\delta m_I = 1$ and $\delta m_I = 2$ transitions, respectively. The echo phase of the NER data is shown in the bottom panel of \textbf{(b)} and \textbf{(c)}.  For the $\delta m_I = 1$ transitions in \textbf{(b)}, the duration of the RF B- and E-field pulses are 20 and 100~$\mu$s, respectively, to generate the similar modulation depth. Both the B- and E-field $\delta m_I = 2$ experiments were performed with the pulse duration of 750~$\mu$s.}
\label{Double_NER}
\end{figure}

We note that quantitatively simulating the NER nutation rates using the nSEC coefficients in Table~\ref{Zgateparameter} is challenging, particularly for the case where $B_0$ is nominally aligned parallel to the crystallographic $c$-axis. In this field orientation, the E-field-driven axial spin Hamiltonian parameters, namely $D$, $A_\parallel$, and $Q_{zz}$, commute with both $\hat{S}_z$ and $\hat{I}_z$. Consequently, to first order, they are not expected to induce nuclear spin transitions when $B_0 \parallel c$-axis. One possible explanation is the presence of a small misalignment of $B_0$. This arises because our experimental apparatus allows only single-axis rotation, making it difficult to achieve the perfect alignment of $B_0$ along the $c$-axis. Such misalignment mixes eigenstates with different $m_s$ and $m_I$ projections, thereby enabling the observation of NERs in the experiment. Though we estimate that any misalignment, calibrated from EPR/ENDOR spectroscopic features, would be small ($<3^\circ$) and not sufficient to explain the nutation rates. Another possibility is that the applied E-field modulates or induces off-diagonal components in the hyperfine ($\bar{\bar{A}}$) and nuclear quadrupole ($\bar{\bar{Q}}$) tensors, such as $A_{xy}$ and $Q_{xz}$, which can efficiently drive nuclear spin transitions~\cite{Asaad2019}. However, the nuclear energy spectrum is relatively insensitive to these off-diagonal terms, making such effects difficult to probe directly using DC E-field measurements.

To further investigate the NERs, we applied $B_0$ perpendicular to the crystal $c$-axis. In this configuration, the electron and nuclear spin eigenstates are predominantly determined by the Zeeman interaction, $B_0(\gamma_S\hat{S}_x + \gamma_I\hat{I}_x)$ for $B_0$ applied along the $x$ direction. Transitions between these Zeeman eigenstates are readily driven by the E-field-modulated axial anisotropy terms. 

The NERs are studied using the pulse sequence shown in Fig.~\ref{Double_NER}(a), in which an additional nuclear spin mixing pulse, driven by either a B-field or an E-field, is incorporated into the ENCT sequence. The frequency of the mixing pulse is chosen to be distinct from that of the other RF B-field pulses. It modifies the nuclear spin coherence by transferring the stored quantum state out of the qubit subspace defined by the two nuclear spin states addressed by the other RF B-field pulses. This sequence is employed instead of that shown in Fig.~\ref{AC_E_Davies}(a) because it provides complete information on both the population and phase of the nuclear spin state, thereby enabling full characterization of the nuclear spin dynamics.

Representative data recorded within the electron spin $\ket{m_s = -1/2}$ manifold are shown in Fig.~\ref{Double_NER}(b) and (c). When the E-field mixing pulse (100~$\mu$s) satisfies the resonance condition for a transition with $\delta m_I = \pm1$, it produces coherent nutations with a rate approximately $\sim20\%$ of that obtained using the corresponding B-field pulse (20~$\mu$s), as shown in Fig.~\ref{Double_NER}(b). By contrast, for a $\delta m_I = \pm2$ transition, no measurable modulation of the echo amplitude is observed when using a B-field mixing pulse, since such transitions are nominally forbidden magnetic-dipole transitions. However, a clear echo modulation is observed upon application of an E-field mixing pulse, with the phase response of the spin echo also agreeing with theoretical expectations (see Supplementary Information for details). This behaviour arises from E-field modulation of the $Q_{zz}\hat{I}_z^2$ interaction, which directly couples states with $\delta m_I = \pm2$ when $\mathbf{B}_0$ is applied perpendicular to the $c$-axis. These results demonstrate that E-field control of spins provides the additional capability of directly coupling non-adjacent eigenstates~\cite{FernndezdeFuentes2024}. Such control can be exploited to replace conventional ladder-type operations, reducing the number of gate operations required for complex quantum algorithms, shortening operation times, and improving overall operational fidelity.

In summary, we demonstrate significant nuclear spin-electric coupling for Mn$^{2+}$ ions doped in ZnO. This coupling originates from the electrically polarisable ZnO host and is further enhanced by the hyperfine-coupled electron spin, enabling efficient and fully electrical implementation of universal single-spin operations ($X$, $Y$, and $Z$ gates). The resulting control efficiencies are comparable to, and in some cases exceed, those achieved using conventional magnetic driving. Importantly, both resonant and non-resonant operations can be realised using uniaxial electric fields applied along the crystallographic $c$-axis. This can significantly simplify device architectures by enabling universal spin control with a single physical gate, representing a key advantage for scalable, all-electrical spin-based nanodevices. Furthermore, the electron spin naturally serves as an ancillary qubit, opening additional pathways for quantum information processing.

Although the nuclear spin coherence time of Mn$^{2+}$ in ZnO is limited by electron spin–lattice relaxation, this limitation can generally be mitigated by operating at sufficiently low temperatures. In addition, decoherence arising from the host nuclear spin bath, primarily due to $^{67}$Zn nuclei with $I = 5/2$ and abundance $\sim4\%$, may be further reduced through isotopic purification~\cite{Pawlis2019}, potentially extending the nuclear spin coherence time $T_{2n}$ into the second regime. Together, these results establish oxides as a promising host material for hyperfine-coupled electron–nuclear spin systems, providing fast, scalable, and electrically driven spin control for next-generation quantum technologies.

\begin{acknowledgments}
This project has received funding from the European Union's Horizon 2020 research and innovation programme under grant agreements 862893 (FATMOLS) and 863098 (SPRING). NF and JL acknowledges support from the Royal Society under grant URF$\backslash$R1$\backslash$201132 and URF$\backslash$R$\backslash$251021. M.V. is grateful to the Hill Foundation for financial support.

\end{acknowledgments}


\bibliography{Nuclear_spin_qudit_electric_control}

\end{document}